\documentclass[usenatbib]{mn2e}
\usepackage{amssymb}
\usepackage{graphicx}
\usepackage{txfonts}

\newcommand{\msun}{{\rm M}_{\sun}}

\title[Viscous propagation of disc variability]
{Viscous propagation of mass flow variability in accretion discs}

\author[A. A. Zdziarski, R. Kawabata and S. Mineshige]
{Andrzej A. Zdziarski,$^1$\thanks{E-mail: aaz@camk.edu.pl (AAZ), kawabata@kusastro.kyoto-u.ac.jp (RK), shm@kusastro.kyoto-u.ac.jp (SM)} Ryoji Kawabata$^2$\footnotemark[1] and Shin Mineshige$^2$\footnotemark[1]\\
$^2$Centrum Astronomiczne im.\ M. Kopernika, Bartycka 18, 00-716 Warszawa, Poland\\
$^1$Department of Astronomy, Kyoto University, Kitashirakawa-Oiwake-cho, Sakyo-ku, Kyoto 606-8502, Japan\\
}

\date{Submitted 2009 July 12; in the original form 2009 February 25}

\pagerange{\pageref{firstpage}--\pageref{lastpage}}
\pubyear{2009}

\begin{document}

\maketitle

\label{firstpage}

\begin{abstract}
We study mass flow rate through a disc resulting from a varying mass supply rate. Variable mass supply rate occurs, e.g., during disc state transitions, and in interacting eccentric binaries. It is, however, damped by the viscosity of the disc. Here, we calculate this damping in detail. We derive an analytical description of the propagation of the flow rate using the solution of Lynden-Bell \& Pringle, in which the disc is assumed to extend to infinity. In particular, we derive the accretion-rate Green's function, and its Fourier transform, which gives the fractional damping at a given variability frequency. We then compare this model to that of a finite disc with the mass supply at its outer edge. We find significant differences with respect to the infinite disc solution, which we find to overestimate the viscous damping. In particular, the asymptotic form of the Green's function is power-law for the infinite disc and exponential for the finite one. We then find a simple fitting form for the latter, and also calculate its Fourier transform. In general, the damping becomes very strong when the viscous time at the outer edge of the disc becomes longer than the modulation time scale. We apply our results to a number of astrophysical systems. We find the effect is much stronger in low-mass X-ray binaries, where the disc size is comparable to that of the Roche lobe, than in high-mass binaries, where the wind-fed disc can have a much smaller size.
\end{abstract}

\section{INTRODUCTION}
\label{intro}

If a binary in which mass is transferred from one component to the other is eccentric, the mass transfer rate will be periodic at the binary period \citep*{afp76,ag80,hlw80,hp84,bb84,bw86,s88,l98,rbm05}. The largest class of such binaries is Be/X-ray binaries, see, e.g., \citet{coe00}, \citet{z02}, \citet{n04} for reviews. All known systems of this type consist of a (strongly magnetized) neutron star orbiting, usually in an eccentric orbit, a high-mass (unevolved) Be star. The accretion occurs via stellar wind and interaction with a circumstellar decretion disc around the massive star, and is maximized around the periastron \citep{w89}. Indeed, most of those systems show X-ray periodicity at their orbital periods, so-called type-I X-ray outbursts \citep*{swr86}, seen, e.g., in A0538--66, EXO 2030+375, A0535+26, AX J0049.4--7323, 4U 0115+63 (see, e.g., references in \citealt{z02}). The accretion onto the neutron star in those systems is thought to occur via a (non-stationary) accretion disc \citep{c86,w89,ho04}. Arguments for the existence of an accretion disc include a lag between the onset of optical activity and an X-ray outburst, abrupt rises and linear decays of outburst (implying the storage of material in a reservoir), and spin-up of the neutron star during outburst \citep{c86}. The last phenomenon is seen during X-ray pulsations at the neutron star spin frequency.

Another class of binaries relevant to this work consists of X-ray pulsars with supergiant donors. Here, accretion is via stellar winds and, in some cases, via a Roche lobe overflow. Some of those systems have eccentric orbits, which then cause periodically varying accretion, e.g., GX 301--2 \citep*{krp80,mp04,lk08}. It has the orbital period of $P_\mathrm{orb} \simeq 41.5$ d and the eccentricity of $e\simeq 0.5$ \citep*{kvn06}. The presence of an accretion disc in this system is discussed by \citet{pg01}, and details of the mass transfer from the companion, including the presence of an accretion stream, are studied by \citet{lk08}.

Another eccentric X-ray binary is the peculiar system Cir X-1, which contains either a low or high-mass donor \citep*{jnb07}, of $P_\mathrm{orb} \simeq 16.5$ d and $e\simeq 0.5$ \citep{p03,w06,jnb07}. The system shows strong periodicity at the orbital period in X-rays \citep*{p03,cco04,w06} and in radio \citep{w77}. Accretion in that system may occur via a Roche-lobe overflow near the periastron \citep{cco04}.

Then there is the case of the ultracompact low-mass X-ray binary 4U 1820--303, where $P_\mathrm{orb}\simeq 685$ s is detected in X-rays \citep*{spw87,z07b}. In addition to the binary period, it shows a strong periodicity at 170 d \citep{w06}, and there is compelling evidence that it is due to the corresponding long-term modulation of the accretion rate \citep*{cg01,z07a}. The main model of the long-term modulation is an interaction with a third star \citep{cg01,z07a}, which causes a periodic modulation of the eccentricity of the inner binary \citep{k62}. Though $e\ll 1$ in this model, the mass supply rate to the outer disc is strongly modulated at both the orbital and superorbital periods. An important issue is whether this orbital modulation of the mass supply rate can be seen in X-rays \citep{ga05,z07b}. 

Enhancement of the mass-transfer rate around the periastron in low-mass X-ray binaries with $e\ll 1$ has also been proposed by \citet{m05} to be responsible for flares discovered from some of X-ray sources in NGC 7469 by \citet*{ssj05}. Again, an issue to examine is whether periodic enhancements in the mass transfer rate in those sources can actually lead to X-ray flares. Viscous damping of a change in the mass supply rate is also an important issue for understanding state transitions in X-ray binaries, see, e.g., \citet{hl02} for the case of a cataclysmic variable.

Thus, the question of the response of an inner part of the disc to a changing mass supply rate in its outer parts is an important astrophysical issue. Here, we study this problem by calculating the effect of the disc viscosity on the propagation of a disturbance from the disc outer edge to the inner one. Our main method (Section \ref{green}) is to calculate Green's functions for the problem, i.e., the disc response to a $\delta$-function pulse of the mass supply. Convolution of the Green's functions with any form of mass supply rate, see equation (\ref{eq:mdotin}), gives the mass accretion rate at the inner edge of the disc. The squared Fourier transform of the Green's function multiplied by the power spectrum spectrum of the mass supply rate gives the power spectrum of the accretion rate, see equation (\ref{eq:convolution}). In particular, if the supply rate is sinusoidal, the absolute value of the Fourier transform gives the factor by which the original amplitude is reduced. After presenting our theoretical results (Sections \ref{green}, \ref{results}), we apply them to X-ray binaries (Section \ref{discussion}).

\section{The Green's functions for the accretion rate and their Fourier transforms}
\label{green}

We calculate propagation of variability of the mass flow rate through an accretion discs by solving time-dependent disc structure. We assume that the disc is geometrically thin, axisymmetric and Keplerian. The equations of continuity and angular momentum transport by viscosity are \citep*{lp74,kfm98}, 
\begin{eqnarray}
\frac{\partial \Sigma}{\partial t} &=& \frac{1}{2 \upi r} \frac{\partial \dot{M}}{\partial r}, \label{eq:con} \\
\dot{M} &=& 6 \upi r^{1/2} \frac{\partial}{\partial r} r^{1/2} \nu \Sigma, \label{eq:ang}
\end{eqnarray}
where $r$, $\Sigma$ and $\dot{M}$ are the disc radial coordinate, the surface density, and the mass flow rate, respectively. Instead of solving the energy equation, we assume that the kinematic viscosity, $\nu$, is a power-law function of only $r$, 
\begin{equation}
\nu = \nu(r_\mathrm{out}) \left( \frac{r}{r_\mathrm{out}} \right)^n,
\label{eq:nu}
\end{equation}
where $r_\mathrm{out}$ is the disc outer radius. In the $\alpha$-viscosity prescription, $\nu=(2/3)\alpha c_\mathrm{s}^2/ \Omega_\mathrm{K}$ (using the conventions of \citealt{kfm98}), where $\alpha$ is the viscosity parameter, $c_\mathrm{s}$ is the sound speed, and $\Omega_\mathrm{K}$ is the Keplerian angular velocity. Then, possible values of $n$ include $1/2$ for a disc without cooling, $3/2$ for an isothermal disc, and $3/5$, $3/4$ for a thin disc supported by gas pressure and dominated by either electron scattering or bound-free Kramers' opacity, respectively \citep{ss73}. In general, the viscosity as a function of radius may not be a power law, e.g., when the dominant source of opacity changes with the radius. We also note that a propagating disturbance in $\Sigma$ will change the local temperature, and thus both the sound speed and viscosity, $\nu$. However, following the treatment of \citet{lp74} and \citet{kfm98}, we neglect here this nonlinearity.

\subsection{The analytical model of an infinite disc}
\label{analytical}

The propagation of variability can be calculated by a simple analytical model assuming the disc extending from infinity to $r=0$, where the density vanishes. At an initial moment, $t=0$, the column density is given by a delta function, $\Sigma=\Sigma_0 \delta (\xi-1)$, where $\xi \equiv (r/r_0)^{1/2}$. Here, $r_0$ is the radius at which mass is supplied to the disc. We will compare in Section \ref{numerical} below the analytical results assuming the disc extending to infinity and the mass injection happening at a finite radius to the numerical ones, where we assume the mass is supplied at the outer edge, $r_{\rm out}$, of a finite disc. With the above assumption for the mass injection, the solution of equations (\ref{eq:con}--\ref{eq:nu}) constitutes the Green's functions for the problem. It can be obtained analytically \citep{lp74,kfm98},
\begin{eqnarray}
\lefteqn{G_\Sigma(r,t) = {\Sigma_0 2|\mu| \xi^{1/\mu-9/2}\over \tau} \exp \left[ -{2\mu^2(\xi^{1/\mu}+1) \over \tau }\right] {\rm I}_{|\mu|} \left[ 4\mu^2\xi^{1/(2\mu)}\over \tau \right] 
\label{eq:sig1},} \\
\lefteqn{G_{\dot{M}}(r,t) =  {\dot M_0 |\mu|\over \tau} \frac{\partial}{\partial \xi} \xi^{1/2} \exp \left[ -{2\mu^2(\xi^{1/\mu}+1) \over \tau }\right] {\rm I}_{|\mu|} \left[ 4\mu^2\xi^{1/(2\mu)}\over \tau \right],} 
\label{eq:mdot1}
\end{eqnarray}
where ${\rm I}_m (z)$ is the modified Bessel function of the first kind,
\begin{equation}
\dot M_0\equiv {4\upi r_0^2\Sigma_0  \over t_{\rm visc}(r_0)},\quad \mu={1\over 4-2n},\quad \tau={t\over t_{\rm visc}(r_0)},\quad t_{\rm visc}={2r^2\over 3\nu},
\label{eq:defs}
\end{equation}
the viscous time scale is defined as $t_{\rm visc}\equiv r/|v_{\rm r}|$ (for a steady-state disc), and $v_r$ is the radial velocity. Hereafter, we will consider the cases with $n<2$, corresponding to $\mu>0$, though the solutions for $n\geq 2$ can be obtained as well \citep{kfm98}. At $n<2$, $t_{\rm visc}$ increases with the increasing $r$, see equation (\ref{eq:defs}). The Green's function for the column density propagation has been used, e.g., by \citet{mw89}, \citet{l97}, \citet{kcg01}, whereas that of equation (\ref{eq:mdot1}) appears not to have been used yet in the literature. 

\begin{figure}
   \centering
\includegraphics[width=8cm,clip]{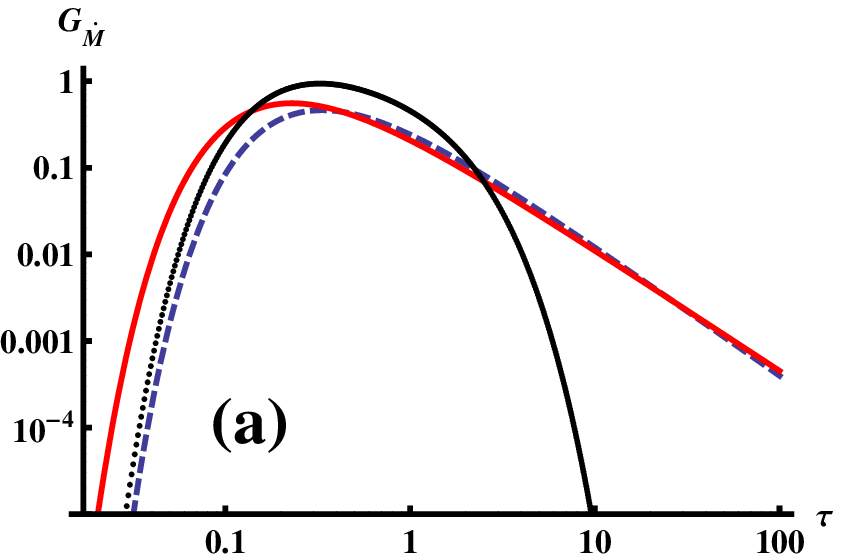}
\includegraphics[width=8cm,clip]{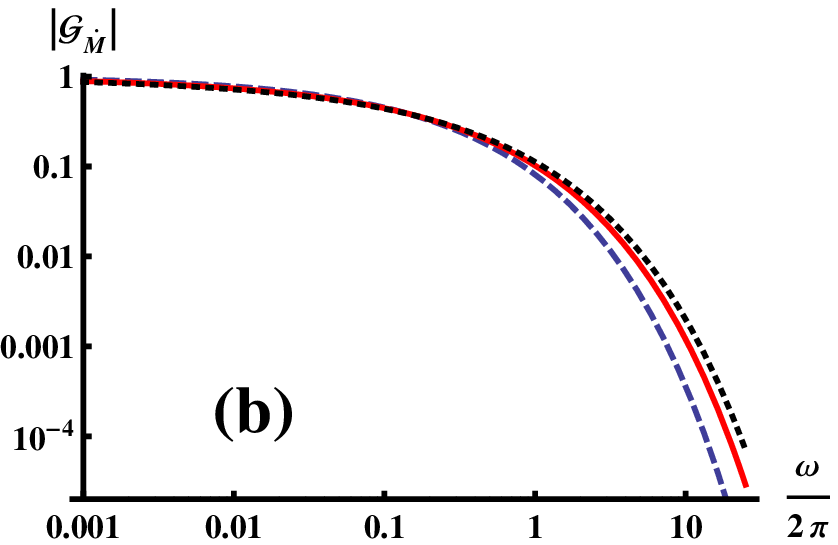}
\caption{(a) The Green's function for the profile of the accretion rate at the inner edge of a disc for an infinite disc with the mass supply at a finite radius, in the cases of $n=3/4$ (red solid curve) and $n=1$ (blue dashes). The black dots (resolved only at small values of $\tau$) show the numerical Green's function for the finite disc at $n=1$, see Section \ref{numerical}. (b) The absolute values of the (infinite-disc) Fourier transforms for $n=3/5$ (black dots), $n=3/4$ (red solid curve) and $n=1$ (blue dashes).
}
\label{f:green}
\end{figure}

Equation (\ref{eq:mdot1}) can be expressed in an explicit form at any radius, see Appendix \ref{radii}. Here we are interested in the accretion rate at the disc inner edge. Hence, we consider the rate at $\xi=0$, which Green's function we denote as $\dot M_0 G_{\dot{M}}(\tau)$, where $G_{\dot{M}}$ is defined to be dimensionless. We find,
\begin{equation}
G_{\dot{M}}(\tau)=
{(2\mu^2)^\mu \over \Gamma(\mu)} \tau^{-1-\mu} \exp\left(-{2\mu^2\over \tau}\right), 
\label{eq:green}
\end{equation}
where $\Gamma$ is the Euler Gamma function, $\int_0^\infty G_{\dot{M}} (\tau){\rm d}\tau=1$, and $G_{\dot{M}} (\tau\leq 0)\equiv 0$. This rate is plotted in Fig.\ \ref{f:green}(a) for $n=3/4$ ($\mu=2/5$) and $n=1$ ($\mu=1/2$).

If the mass supply rate at $r_0$ is $\dot M_{\rm out}(\tau)$, the rate at the inner disc edge is a convolution,
\begin{equation}
\dot M_{\rm in}(\tau)= \int_0^\infty \dot M_{\rm out}(\tau-\tau_1)  G_{\dot{M}}(\tau_1) {\rm d}\tau_1.
\label{eq:mdotin}
\end{equation}
The Fourier transform of $\dot M_{\rm in}(t)$ is then the product of the individual transforms, i.e., 
\begin{equation}
\dot{\cal M}_{\rm in}(\omega)=\dot{\cal M}_{\rm out}(\omega) {\cal G}_{\dot{M}} (\omega).
\label{eq:convolution}
\end{equation}
The Fourier transform, $\int_{-\infty}^\infty G_{\dot{M}}(\tau) {\rm e}^{-{\rm i}\omega \tau} {\rm d}\tau$, of the Green's function of equation (\ref{eq:green}) is,
\begin{eqnarray}
\lefteqn{{\cal G}_{\dot{M}}(\omega) = { 2 (2\mu^2 \omega)^{\mu/2} \over \Gamma(\mu)} {\rm e}^{{\rm i}\upi\mu/4} {\rm K}_\mu\left[2 (1+{\rm i})\mu \sqrt{\omega}\right],
\label{eq:fourier} }\\
\lefteqn{{\cal G}_{\dot{M}}(0)=1,\quad \left|{\cal G}_{\dot{M}}(\omega\gg 1)\right|\rightarrow {\sqrt{\upi}\over \Gamma(\mu)}(2\mu^2\omega)^{(2\mu-1)/4} {\rm e}^{-2\mu \sqrt{\omega}},
\label{eq:fourier_asymptotic} }
\end{eqnarray}
where ${\rm K}_m (z)$ is the modified Bessel function of the second kind, $\omega=2\upi f t_{\rm visc}(r_0)$, and $f$ is the frequency. For $n=1$ ($\mu=1/2$), the transform assumes a particularly simple form,
\begin{equation}
{\cal G}_{\dot{M}}(\omega) =  {\rm e}^{-(1+{\rm i})\sqrt{\omega}} ,\qquad \left|{\cal G}_{\dot{M}}(\omega)\right| =  {\rm e}^{-\sqrt{\omega}}.
\label{eq:fourier1}
\end{equation}
Fig.\ \ref{f:green}(b) shows $|{\cal G}_{\dot{M}}(\omega)|$. If the original signal is sinusoidal, its amplitude is reduced by $\left|{\cal G}_{\dot{M}} (\omega)\right|$. We see that the viscous damping increases with the increasing $n$.  This is caused by the viscous time within the disc, $t_\mathrm{visc}(r) = t_\mathrm{visc} (r_\mathrm{out}) (r/r_\mathrm{out})^{2-n}$, being longer for a larger $n$ at a given $t_\mathrm{visc}(r_\mathrm{out})$.

In order to account for the effect of viscous diffusion on the power spectrum of an arbitrary mass supply rate, we need to multiply its power spectrum by $\left|{\cal G}_{\dot{M}}(\omega)\right|^2$. In general, damping (time-dependent or stationary) of a variable component of the accretion rate due to the viscous transport through the disc can be calculated using the convolution of the light curves, equation (\ref{eq:mdotin}). Damping can also be calculated by integration of resulting power spectra, equation (\ref{eq:convolution}). Using the former method to calculate the time evolution for a step-function mass-supply either increase from 0 to $\dot M_0$ or decrease from $\dot M_0$ to 0 at $\tau=0$ yields at $\tau\geq 0$,
\begin{equation}
\dot M_{\rm in}(\tau)=\dot M_0 f(\tau),\quad \dot M_{\rm in}(\tau)=\dot M_0 [1-f(\tau)],
\label{eq:step}
\end{equation}
respectively, where
\begin{equation}
f(\tau)= \cases{
\Gamma(\mu,2\mu^2/\tau)/ \Gamma(\mu), &$\mu>0$;\cr
\exp(-2/\tau), &$\mu=1\,(n=3/2)$,\cr}
\label{eq:step_f}
\end{equation}
and $\Gamma(x,y)$ is the incomplete Gamma function. By subtraction, we can also calculate $\dot M_{\rm in}$ for a rectangular pulse of $\dot M_{\rm out}$. Note that our results do not depend on the normalization of $\dot{M}$ as the equations are linear. In particular, a constant $\dot M$ can be added to any of the solutions.

\subsection{Finite disc}
\label{numerical}

In our numerical calculations, we assume the disc is cut off at $r_{\rm out}$, at which the mass accretion rate $\dot{M}_{\rm out}(t)$ is given as a boundary condition, and its density is zero at the inner edge, $r_{\rm in}$. We also assume conservation of angular momentum at $r_{\rm out}$, and do not allow a disc formation beyond it. Since the kinematic viscosity is assumed to be a power-law function of $r$, the results depend only on the ratio $r_{\rm out}/r_{\rm in}$. We calculate the time evolution of the disc structure by solving equations (\ref{eq:con}--\ref{eq:nu}). We divide the disc radius into $N=10^2$--$10^3$ grid points. 

\begin{figure}
   \centering
\includegraphics[width=8cm,clip]{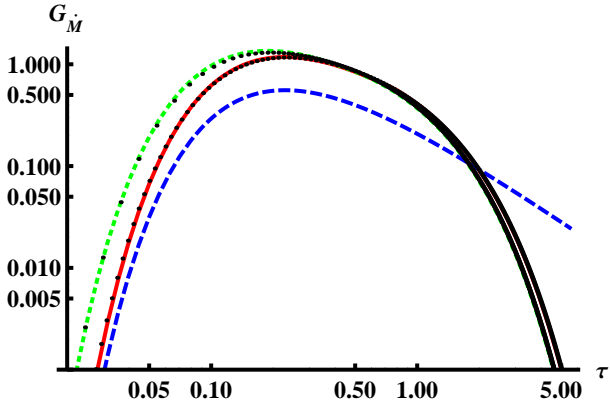}
\caption{The Green's functions for the profile of the accretion rate at the inner edge of a finite disc with the mass supply at the outer edge. The black dots (resolved only at small values of $\tau$) give the numerical solution, and the green dots and the red solid curve give the fit of equation (\ref{eq:fit})  for $n=3/5$ and 3/4, respectively. The results are compared to those for the infitite disc at $n=3/4$, shown by the blue dashes. 
}
\label{f:green_num}
\end{figure}

In order to calculate Green's function for this problem, we consider a pulse with the duration of $10^{-5} t_{\rm visc}(r_{\rm out})$ at the outer edge of a disc with $r_{\rm out}/r_{\rm in}=10^5$, for $n=3/4$, 3/5 and 1. We now define $\tau\equiv t/t_{\rm visc}(r_{\rm out})$. We have found out that the outer boundary condition changed with respect to that in the analytical model results in a rather significant change of the shape of the Green's function. Namely, whereas the analytical Green's function has a power-law asymptotic (at $\tau\gg 1$) form, $G_{\dot{M}}(\tau) \propto \tau^{-1-\mu}$, equation (\ref{eq:green}), the numerical Green's function behaves like $G_{\dot{M}}(\tau) \propto \exp(-\beta \tau)$, where we have found the dependence,
\begin{equation}
\beta= {9\over 4}-n= {1\over 4}+{1\over 2\mu}.
\label{eq:beta}
\end{equation}
On the other hand, the shape (but not the normalization) of the finite-disc Green's function at low $\tau$ is found to be virtually identical to that of equation (\ref{eq:green}). Combining these two dependencies, we find excellent fits to our numerical results by,
\begin{equation}
G_{\dot{M}}(\tau)\simeq N\cases{{(2\mu^2)^\mu \over \Gamma(\mu)} \tau^{-1-\mu} \exp\left(-{2\mu^2\over \tau}\right), &$\tau<\tau_0$;\cr
\eta \exp\left[-\left({1\over 4}+{1\over 2\mu}\right)\tau\right], &$\tau\geq \tau_0$.\cr}
\label{eq:fit}
\end{equation}
Here, the only free fitting parameter is $\tau_0$. The value of $\eta$ follows from the continuity at $\tau_0$, and $N$ follows from the integral over $\tau$ being unity. Both conditions are imposed during $\chi^2$ fitting (of $\ln G_{\dot{M}}$). For completeness, we give here all the obtained values, $\tau_0=0.781$, 0.400, 0.301, $N=2.190$, 2.139, 2.296, $\eta =0.901$, 0,844, 0.669, for $n=3/5$, 3/4, 1, respectively. We see that the initial (at low $\tau$) Green's function is larger by $N\!\sim\! 2$ than that for the infinite disc. This can be explained by the propagation of the same $\dot M$ being now only inward. The Green's functions and their fits for $n=3/5$ and 3/4 are plotted in Fig.\ \ref{f:green_num}. We see that equation (\ref{eq:fit}) provides excellent fits to the shown numerical results, within $\la 4$ per cent at $\tau\ga 0.05$. On the other hand, the shape of equation (\ref{eq:green}) gives a slightly worse description of the Green's function at $\tau<\tau_0$ for $n=1$, which value of $n$, however, does not correspond to any physical disc. Figs.\ \ref{f:green}(a) and \ref{f:green_num} compare the finite-disc Green's function to those for the infinite disc, equation (\ref{eq:green}). 

\begin{figure}
   \centering
\includegraphics[width=8cm,clip]{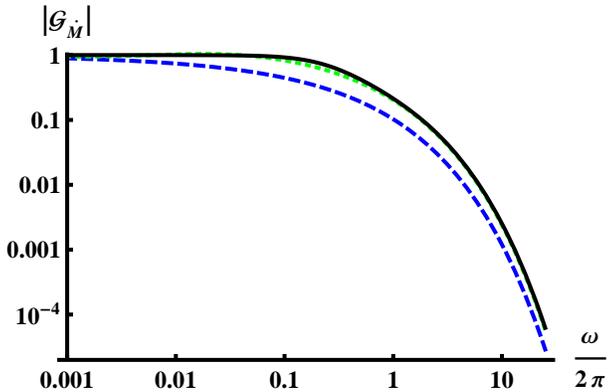}
\caption{Comparison of the absolute values of the Fourier transform and its approximation for $n=3/4$. The black solid curve gives both the transform of the numerical Green's function for the finite disc and the transform of equation (\ref{eq:fit}), which are indistinguishable from each other on this plot. The green dots give the approximation of equation (\ref{eq:correction}). The blue dashes give the analytical result for the infinite disc, equation (\ref{eq:fourier}).
}
\label{f:fourier_num}
\end{figure}

The Fourier transform of the above Green's functions, see Fig.\ \ref{f:fourier_num}, can be obtained either directly from the numerical results or by integration of equation  (\ref{eq:fit}). The Green's function and its transform obtained here for a finite disc can be used analogously to those for the infinite disc, Section \ref{analytical}, e.g., to solve a time-dependent disc behaviour, damping of a periodic variability, or the effect on power spectra. 

The narrowness of the Green's functions for the finite disc results in the transform for the finite disc being larger by $\sim\!2$ at high $f$ than that for the infinite disc, though both equal 1 at $f\rightarrow 0$, see Fig.\ \ref{f:fourier_num}. We have found that the ratio of damping factor for the finite disc to that for the infinite disc, equation (\ref{eq:fourier}), is roughly given by a phenomenological factor,
\begin{equation}
{\left|{\cal G}_{\dot{M}}(\omega)\right|_{\rm finite} \over \left|{\cal G}_{\dot{M}}(\omega)\right|_{\rm infinite}}\sim {1+20\omega\over 1+10\omega}.
\label{eq:correction}
\end{equation}
This also provides a good fit to the numerical results [though worse than that using the Fourier transform of equation (\ref{eq:fit})], as shown in Fig.\ \ref{f:fourier_num}.

\begin{figure}
   \centering
\includegraphics[width=8cm,clip]{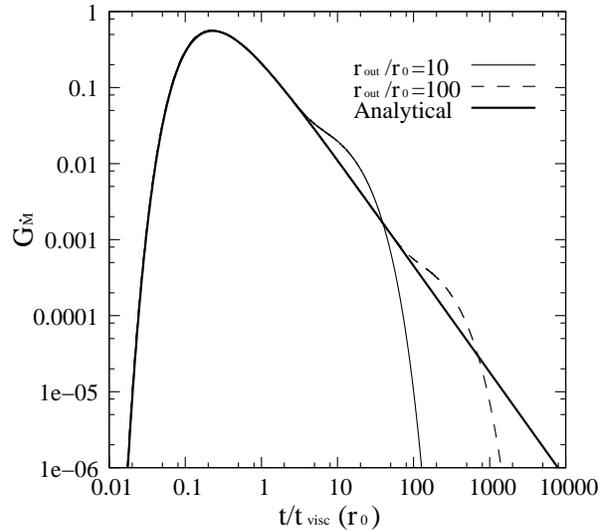} 
\caption{The Green's function for the profile of the accretion rate at the inner edge of a finite disc with the mass supply at $r_0 \ll r_{\rm out}$ for $n=3/4$. The results are compared to those for an infinite disc, equation (\ref{eq:green}), shown by the thick solid curve. }
\label{f:pulse}
\end{figure}

In order to explain the rather significant difference between the form of the Green's function between the finite and infinite disc, power law vs.\ exponential, we have solved numerically the case of mass supply at an intermediate radius, $r_0 \ll r_{\rm out}$, where $r_{\rm out}$ is the outer radius of a finite disc. We have assumed $r_{\rm out}/r_{\rm in}=10^5$ and $n=3/4$. We have found that the exponential behaviour of the response to a pulse is due to the finite size of the disc. The Green's function first follows the form of that for the infinite disc, with a power law decay, but it starts to depart from the infinite-disc solution at $t/t_{\rm visc}(r_0)\sim r_{\rm out}/r_0$ and then it decays exponentially. This is illustrated in Fig.\ \ref{f:pulse}, for two choices of $r_{\rm out}/r_0$. The-power law asymptotic form is due to the assumption of the infinite $r_{\rm out}$, allowing the mass flow to diffuse back to the inner radius back from arbitrarily large radii.

We also note that the overall form of the finite-disc Green's function, not only its exponential decay, appears common for diffusion problems. For example, the Green's function for Thomson-scattering diffusion of photons out of a spherical electron cloud has a very similar shape to that found for the present problem, see, e.g., \citet*{z09}. 

\section{Further numerical models}
\label{results}

\begin{figure}
   \centering
\includegraphics[width=8cm,clip]{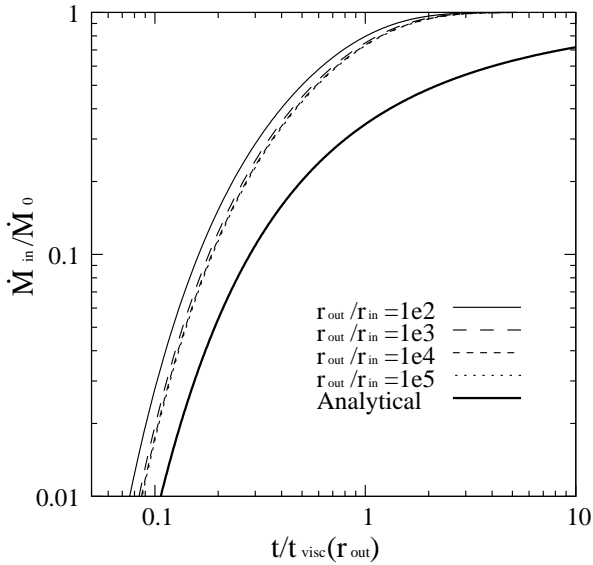} 
\caption{Time evolution of the mass accretion rate for a step-function increase of the mass supply rate at the disc outer edge from $\dot{M}_{\rm out}=0$ up to $1$ at $t = 0$, for $r_{\rm out}/r_{\rm in}=10^2$--$10^5$. These results are compared to those for the infinite disc, equation (\ref{eq:step}), shown by the thick solid curve.}
\label{f:increase}
\end{figure}

We present most of numerical results in this section for $n=3/4$, appropriate for outer disc regions, where most of the viscous damping takes place. We assume $\dot M_{\rm out} = \dot M_{\rm in}$ at $t=0$, and we start to change $\dot M_{\rm out}$ at $t>0$. We consider first the mass supply rate increasing as a step function from $\dot M_{\rm out}=0$ to 1. Fig.\ \ref{f:increase} shows the resulting evolution of the mass accretion rate. The inner mass accretion rate increases on the time scale of $\sim t_\mathrm{visc} (r_\mathrm{out})$. The steady state is reached after a few $t_\mathrm{visc} (r_\mathrm{out})$. The accretion rate reaches an asymptotic temporal profile independent of $r_{\rm out}/r_{\rm in}$ at its very large values, which is due to most of the viscous damping taking place in outer parts of the disc. We find the steady state is achieved much faster than in the case of an infinite disc, which is due to its Green's function being much broader than the one for finite disc, see Figs.\ \ref{f:green}(a), \ref{f:green_num}.

Then, we consider periodic modulation of the mass supply rate. After turning on the modulation, the disc settles down into a steady-state periodic behaviour after some number of $t_\mathrm{visc} (r_\mathrm{out})$. We define the steady-state fractional variability amplitude, 
\begin{equation}
A(r)= \frac{\dot{M}_\mathrm{max}(r)- \dot{M}_\mathrm{min}(r)}{\langle \dot{M}\rangle},
\label{eq:amplitude}
\end{equation} 
where $\dot{M}_{\rm max}(r)$ and $\dot{M}_{\rm min}(r)$ are the maximum and minimum flow rates, respectively, and $\langle \dot M\rangle$ is the average rate.

\begin{figure}
   \centering
\includegraphics[width=8cm,clip]{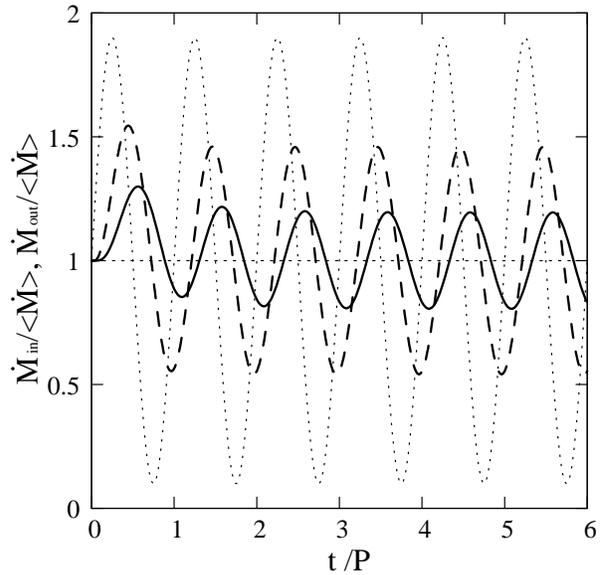} 
\caption{Time evolution of the mass accretion rate at $n=3/4$ and $r_{\rm out}/r_{\rm in}=10^4$ for the case of the sinusoidal modulation with $A(r_{\rm out})=1.8$, shown by the dotted curve, turned on at $t=0$. The solid and dashed curves show $\dot{M}_{\rm in}(t)$ for $t_{\rm visc}(r_{\rm out})=P$ and $0.4 P$, respectively. }
\label{f:mdot}
\end{figure}

\begin{figure}
   \centering
\includegraphics[width=8cm,clip]{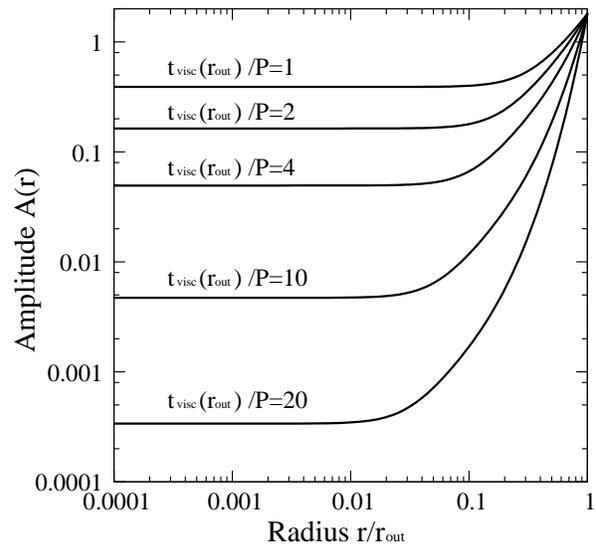} 
\caption{The amplitude of the steady-state periodic variability of the mass flow rate as a function of the radius, $A(r)$, for a number of values of $t_{\rm visc}(r_{\rm out})/P$. The parameters are $n=3/4$, $r_{\rm out}/r_{\rm in}=10^4$ and the amplitude of the sinusoidal modulation is $A(r_{\rm out})=1.8$.}
\label{f:radial}
\end{figure}

We first consider a sinusoidal modulation, 
\begin{equation}
\dot{M}_\mathrm{out} = \langle \dot{M}\rangle \left[ 1+{A(r_{\rm out})\over 2} \sin \frac{2 \upi t}P \right],
\label{eq:sinus}
\end{equation}
where $P$ is the modulation period (usually, but not necessarily, equal to the binary period, $P_{\rm orb}$). Fig.\ \ref{f:mdot} shows an example of the disc settling into a periodic behaviour after starting the modulation at $t=0$.
The inner mass accretion rate shows regular sinusoidal behaviour after several $t_{\rm visc}(r_{\rm out})$. Fig.\ \ref{f:radial} shows the fractional variability as a function of the radius for this case for several values of $t_{\rm visc}(r_{\rm out})/P$. We see that the final value of the modulation amplitude, $A_{\rm in}$, is reached already at $r\gg r_{\rm in}$. Obviously, the viscous damping becomes stronger with the increasing $t_{\rm visc}(r_{\rm out})/P$.

\begin{figure}
   \centering
\includegraphics[width=8cm,clip]{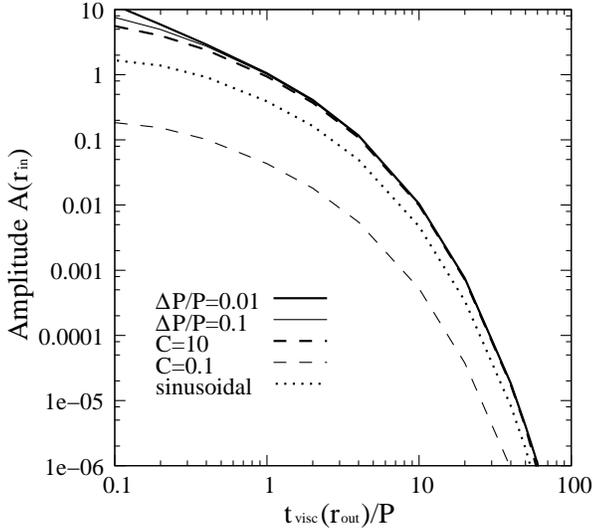} 
\caption{The amplitude of the periodic variability of the mass accretion rate at the inner edge of the disc for $\Delta P/P=0.01$ (thick solid curve) and $0.1$(thin solid curve), for $C = 10$ (thick dashes) and $1$ (thin dashes), and for sinusoidal case with $A(r_{\rm out})=1.8$ (dots). In all cases, $n=3/4$ and $r_{\rm out}/r_{\rm in}=10^4$.}
\label{f:tvis}
\end{figure}

In the case of a large eccentricity, accretion can take place only around periastron, as it is the case, e.g., in Be X-ray binaries. We thus approximate the disc supply rate as a periodic rectangular function,
\begin{equation}
\dot{M}_\mathrm{out} =  \langle \dot{M}\rangle\cases{                                                 P/ \Delta P, & $k < t/P < k + \Delta P/P$;\cr
0,  & otherwise,\cr}
\label{rectangular}
\end{equation}
where $\Delta P$ is the duration of the accretion, and $k$ is an integer. We also consider $\dot{M}_\mathrm{out}$ corresponding to a Roche-lobe overflow,
\begin{equation}
\dot{M}_\mathrm{out} = \dot M_1 \exp \left( C \sin \frac{2 \upi t}{P} \right), 
\label{roche}
\end{equation}
see, e.g., \citet{z07a}, who obtained $C\simeq 1.5$ in the case of the ultracompact low-mass X-ray binary 4U 1820--303 at the maximum of its modelled eccentricity. Here $\dot M_1$ is a normalization factor. Fig.\ \ref{f:tvis} shows the relative variability amplitude at the inner edge as a function of $t_{\rm visc}(r_{\rm out})/P$ for sinusoidal, periodic rectangular and Roche-lobe overflow cases. When $t_\mathrm{visc} (r_\mathrm{out}) \ga 10 \ P$, the amplitude is very strongly damped and $A(r_\mathrm{in})<0.01$ even if $P_{\rm orb}/\Delta P$ or $C$ are very large. The form of the damping is qualitatively similar in all cases. Thus, our result that the amplitude is strongly damped for a large $t_\mathrm{visc}(r_\mathrm{out}) /P$ does not depend on the functional form of the mass supply rate.

\section{Applications to X-ray binaries}
\label{discussion}

The viscous time is, in the $\alpha$ prescription, given by 
\begin{equation}
t_\mathrm{visc}= {1\over \alpha\Omega_\mathrm{K}} \left(c_\mathrm{s}\over v_\mathrm{K}\right)^{-2},
\label{eq:tvisc}
\end{equation}
where $v_\mathrm{K}$ is the Keplerian velocity (and $c_\mathrm{s} / v_\mathrm{K} = H/r$, where $H$ is the disc scale height). The period of the variability of the mass supply rate, $P$, usually equals the orbital period, which is given by the Kepler law, $P_\mathrm{orb}=2 \upi \sqrt{a^3/ GM_1(1+q)}$, where $q=M_2/M_1$, and $M_1$, $M_2$ and $a$ is the mass of the compact star, the donor mass, and the semi-major axis, respectively. Then (see also \citealt{ga05}),
\begin{equation}
\frac{t_\mathrm{visc}(r_\mathrm{out})}{P_\mathrm{orb}} =  {(1+q)^{1/2}\over 2 \upi \alpha} \left( \frac{c_\mathrm{s}}{v_\mathrm{K}} \right)^{-2} \left( \frac{r_\mathrm{out}}{a} \right)^{3/2}.
\label{eq:ratio}
\end{equation}
In the standard thin accretion disc, the ratio $c_\mathrm{s}/v_\mathrm{K}$ is very small, $\sim 0.02$. In low-mass X-ray binaries, the size of the disc is comparable to the size of the Roche lobe around the compact object, $R_1$ \citep{pp77,ga05}, which, for $e\ll 1$ and $q\la 1$ characteristic to that class of system, is in turn comparable to $a$. Thus, $r_\mathrm{out}/a\sim 0.5$, and then the viscous time is much longer than the orbital period, $t_\mathrm{visc} (r_\mathrm{out}) /P_\mathrm{orb} \ga 10^2/\alpha$. Then, our results imply that any periodic variability of the mass accretion rate at the inner disc is completely negligible in the standard model even if the mass supply to the outer disc is periodic. This implies, in particular, that the $\sim 1$ per cent orbital modulation seen in X-rays in 4U 1820--303 is not due to periodicity of the mass supply predicted by the triple model of that system \citep{z07a}, provided the accretion disc is standard. This confirms the supposition of \citet{z07b}. On the other hand, the viscous damping of the superorbital variability in 4U 1820--303, with $P\simeq 170$ d, is negligible unless $\alpha\la 0.01$. 

We also note that our results rule out the eccentric low-mass X-ray binary model of \citet{m05} for the X-ray flares in NGC 7469 found by \citet{ssj05}. \citet{m05} claims that since the flares are super-Eddington, the accretion discs are radiation-pressure dominated and then $c_{\rm s}\sim v_{\rm K}$ and $t_{\rm visc}< P_\mathrm{orb}$. However, super-Eddington discs are radiation-pressure dominated only in inner regions whereas the outer parts are still gas-pressure dominated and geometrically thin (unless the accretion rate is extremely high). Specifically, for parameters suggested by \citet{m05}, $M_1=1.4\msun$, $M_2=1\msun$, $P_{\rm orb}=15$ hr, $\alpha=0.3$, the disc size of 0.9 of the Roche lobe size, $\dot M c^2/L_{\rm E}=100$ (where $L_{\rm E}$ is the Eddington luminosity), we have obtained $t_{\rm visc}\sim 10^2 P_{\rm orb}$ using standard disc equations. Obviously, there will be then no observable enhancement of the X-ray flux around the periastron.

We also note that \citet{s09} have found no periodicities in X-ray flares in NGC 7469, which provides independent evidence against the eccentric binary model. The flares may instead have similar nature to the aperiodic flares discovered from the black hole binary Cyg X-1 with a range of time scales \citep{g03,gz03}. 

\citet{ga05} found breaks in the power spectra in a number of low-mass X-ray binaries at frequencies proportional to the orbital frequencies. They interpreted those frequencies as equal to $t_{\rm visc}^{-1}$, which then implied that $t_{\rm visc}$ were much shorter than that predicted by the standard model, which effect they explained by the presence of hot coronal flows covering outer parts of the disc. In particular, they claimed $t_{\rm visc}\sim 300\, {\rm s}\sim P_\mathrm{orb}/2$ in 4U 1820--303. After such a short time interval, most of the large, $\ga 10$, mass-supply modulation at the outer disc due to eccentricity predicted by the triple model \citep{z07a} would still be present close to the neutron star surface. However, such modulation is clearly not seen. Thus, $t_{\rm visc}\sim 300$ s claimed by \citet{ga05} for 4U 1820--303 is ruled out in the framework of this model. On the other hand, it may be that the cold disc is covered by a fast coronal flow having a short viscous time and carrying $\sim$1 per cent of the mass flow, which could then account for the observed X-ray periodicity.

Then, high-mass X-ray binaries have usually $q\gg 1$, for which the size of the Roche lobe of the compact object in the circular case is \citep{p71},
\begin{equation}
{R_1\over a}= {2\over 3^{4/3} (1+q)^{1/3}}.
\label{eq:roche}
\end{equation}
If $r_\mathrm{out}$ were $\sim R_1$ (as expected for Roche lobe overflow), the time scale ratio of equation (\ref{eq:ratio}) would become $\propto q^0$, and the conclusion of $t_\mathrm{visc} (r_\mathrm{out}) /P_\mathrm{orb}\gg 1$ would still hold. Taking into account eccentricity can be done in a simplified way by a substitution of  $a\rightarrow a(1-e)$ (i.e., using the periastron separation as constraining the disc size), which reduces the time scale ratio by $(1-e)^{3/2}$, which is significant only for $e\sim 1$.

However, discs in wind-accreting systems may be truncated at radii much smaller than the Roche lobe radii of the compact object, e.g., \citet{sl76}. The outer edge of the disc is then determined by the specific angular momentum in a Keplerian orbit around the compact object, $(G M_1 r_{\rm out})^{1/2}$, being equal to the angular momentum carried by the accreting gas, $(1/2)v_{\rm orb} r_{\rm a}^2/a$, where $r_{\rm a}\simeq 2 GM_1/v_{\rm rel}^2$ is the accretion radius, $v_{\rm orb}$ is the relative velocity of the stars, $v_{\rm rel}$ is the relative velocity of the compact object and the wind, $v_{\rm rel}^2=v_{\rm orb}^2+v_{\rm wind}^2$, and $v_{\rm wind}$ is the wind velocity \citep{sl76}. This yields, 
\begin{equation}
{r_{\rm out}\over a}= {4d^3\over b^8 (1+q)^3 (2a- d)^3},
\label{eq:wind}
\end{equation}
where $d$ is the orbital-phase dependent distance between the stars and $b=v_{\rm rel}/v_{\rm orb}$. Note that equation (\ref{eq:wind}) can be used only if it yields $r_{\rm out}<R_1$. \citet{pg01} assumed that the neutron star in GX 301--2 accretes from a slow, dense, circumstellar wind, in which case $b\simeq 1$. Then at periastron, $r_{\rm out}/ a$ becomes $4[(1-e)/ (1+q)(1+e)]^3$, which for $q\simeq 20$, $e\simeq 0.5$ \citep{kvn06} is $\simeq 2\times 10^{-5}$. Equation (\ref{eq:ratio}) then yields $t_\mathrm{visc}(r_\mathrm{out})/ P_\mathrm{orb} \sim 2\times 10^{-4}/\alpha$. Thus, the viscosity damping of the orbital periodicity is negligible in GX 301--2. Similar considerations apply to eccentric Be/X-ray binaries.

The case of Cir X-1 is less clear given the uncertainty about its system parameters. \citet{cco04} have modelled the orbital phase dependence of the mass transfer rate in Cir X-1 using the model of \citet{bb84}, and they estimated $t_{\rm visc}\sim 10$ d, i.e., also less than its $P_{\rm orb}$. Thus, viscosity has probably a minor effect on the observed orbital periodicity in that object. Observationally, we do see strong periodicity at $P_{\rm orb}$, which also implies $t_{\rm visc}< P_{\rm orb}$.

We also note that the time scale of state transitions in black-hole or neutron-star binaries is likely to be the viscous time scale at the radius of the mass supply (different from the case of outbursts of either X-ray or dwarf novae, where the time scale is by a factor of $\sim\! H/r$ shorter, \citealt{m84,kfm98}). Then, our results appear to rule out the mass supply in Cyg X-1 (with $P_{\rm orb}=5.6$ d) being at the outer edge of a standard disc with the size comparable to its Roche lobe radius. If this were the case, $t_{\rm vis}$ would be $\sim\! 10^2 P_{\rm orb}/\alpha$. Then, any change of the mass supply rate to the disc would result in a change of the mass accretion rate in the inner disc being on a time scale of $\gg P_{\rm orb}$. This is much longer than the observed time scale of state transitions in this system of a few $P_{\rm orb}$ (e.g., fig.\ 5 in \citealt{zg04}). Thus, the state transitions in Cyg X-1 cannot be caused by a change of $\dot M$ at the outer edge of a standard accretion disc. Indeed, though the donor almost fills its Roche lobe, the wind accretion is still dominant \citep{g03}, and the size of the disc is then likely to be much smaller than that of the Roche lobe of the black hole. Equation (\ref{eq:wind}) with $q\simeq 2.8\pm 0.4$, $b\sim 2$ \citep{g03} and $e=0$ yields then a rather small disc, with $r_{\rm out}/ a\simeq 3\times 10^{-4}$, at which equation (\ref{eq:ratio}) yields $t_\mathrm{visc} (r_\mathrm{out})/ P_\mathrm{orb}\sim 0.004/\alpha$. (Note that the wind velocity around the compact object in Cyg X-1 is not well determined, which makes the determination of $r_{\rm out}/ a$ rather uncertain.) The combination of the actual disc size in Cyg X-1 (which may be intermediate between those of the pure Roche-lobe overflow and wind accretion cases) and the viscosity parameter may correspond to $t_\mathrm{visc} (r_\mathrm{out})\sim P_\mathrm{orb}$, as it appears to be implied by the observations. Still, a large accretion disc can be present if the actual viscous time is much shorter than the standard one, as in the model of \citet{ga05}.

We can also invert equation (\ref{eq:ratio}) in order to obtain the disc size corresponding to a given $t_{\rm visc}(r_{\rm out})/P_{\rm orb}$. This yields,
\begin{equation}
\frac{r_\mathrm{out}}{a}= {(2 \upi \alpha)^{2/3} \over (1+q)^{1/3}} \left( c_\mathrm{s}\over v_\mathrm{K} \right)^{4/3} \left[ t_\mathrm{visc}(r_\mathrm{out})\over P_\mathrm{orb}\right]^{2/3}.
\label{eq:radius}
\end{equation}
We see that $t_{\rm visc}(r_{\rm out})\sim P_{\rm orb}$ implies $r_{\rm out}/a\ll 1$ in general. 

Throughout the present study, we have assumed that a disc has an axisymmetric structure and have examined only the viscous damping effect. On the other hand, \citet{ho05} performed three-dimensional smoothed particle hydrodynamics simulations of Be/X-ray systems and claimed that time-dependent mass transfer to the disc will produce a non-axisymmetric structure because of the ram pressure by the incoming stream. Then, a spiral wave is created and its inward propagation significantly enhances the mass-accretion rate onto the neutron star (see also \citealt*{hmh08} in the context of binary black holes). If this is the case, it is possible to produce appreciable light variations even when the viscous time scale at the outer edge of the disc is much longer than the time scale of the mass input. Our simple analyses presented in this paper does not allow to examine this effect and multi-dimensional simulations are necessary in future work.

In our applications, we have considered only binaries. However, our results on the viscous damping of variability hold for any accretion disc, also those around supermassive black holes in active galactic nuclei.

\section{Conclusions}
\label{conclusions}

We have calculated in detail the effect of viscous damping of variability in accretion discs. We have first considered the infinite-disc model of \citet{lp74}. For it, we have obtained a simple form of the Green's function for $\dot M$, equation (\ref{eq:green}), as well as its Fourier transform, equation (\ref{eq:fourier}). The latter gives the fractional viscous damping at a given variability frequency, approximately given by $\sim\! \exp(-\sqrt{\omega})$. As a simple application of our Green's function, the case of a step function increase of $\dot M$ yields an approximately exponential approach to the steady state, $\sim\! \exp(-1/\tau)$. 

We have then calculated numerically a realistic finite-disc model with the mass supply at its outer boundary. We have found significant differences with respect to the infinite-disc model. In particular, a single pulse at the disc outer edge results in an exponential decay after a fraction of the viscous time at the disc outer edge, whereas the infinite-disc model yields a much broader power-law decay. The power-law behaviour is due to the infinite size, allowing the flow to diffuse back from arbitrarily large radii. Given the differences, we have fitted the finite-disc Green's functions by a simple form, and calculated its Fourier transform. We have also calculated numerically time-dependent disc behaviour for a number of more complex cases, in particular for periodic modulations of the mass supply rate.

We have applied our results to a number of X-ray binaries. Periodic variability resulting from eccentricity is usually very strongly damped in low-mass X-ray binaries, which accretion discs have the size comparable to that of the Roche lobe of the compact object. This is the case, in particular, in 4U 1820--303 or in flaring binaries in NGC 7469. 

The periodic mass transfer in eccentric binaries is much easier to observe in high-mass X-ray binaries, where accretion is via a wind and the accretion discs can be truncated at radii much smaller than the size of the Roche lobe, resulting in the viscous time being relatively short. In particular, the viscous damping is negligible in GX 301--2, Be/X-ray binaries and Cir X-1. In the case of Cyg X-1, the observed time scale at which state transitions take place is comparable to its orbital period, which implies the accretion disc being also truncated at a radius much lower than the Roche-lobe size. This may be due to the presence of wind accretion in this system.

\section*{ACKNOWLEDGMENTS}

We thank M. Gilfanov, J.-P.\ Lasota and R. Misra for valuable discussions, and the referee for valuable suggestions. This research has been supported in part by the Polish MNiSW grants NN203065933 and 362/1/N-INTEGRAL/2008/09/0, and the Polish Astroparticle Network 621/E-78/BWSN-0068/2008.

\appendix
\section{Viscous propagation of $\dot{\mathbfit M}$ between two disc radii}
\label{radii}

The explicit form of $G_{\dot{M}}(r,t)$ of equation (\ref{eq:mdot1}) for $\mu>0$ is,
\begin{eqnarray}
\lefteqn{G_{\dot{M}}(\xi,\tau) =  {\dot M_0 2\mu^2\xi^{(1/\mu-1)/2} \over \tau^2} \exp \left[ -{2\mu^2(\xi^{1/\mu}+1) \over \tau}  \right] \nonumber}\\
\lefteqn{\quad\times
\left\{ {\rm I}_{\mu-1} \left[ 4\mu^2\xi^{1/(2\mu)}\over \tau \right]-\xi^{1/(2\mu)}{\rm I}_{\mu} \left[ 4\mu^2\xi^{1/(2\mu)}\over \tau \right]\right\}. }
\label{eq:md_general}
\end{eqnarray}
At $n=1$ ($\mu=1/2$),
\begin{equation}
G_{\dot{M}}(\xi,\tau) =  {\dot M_0 \over \sqrt{2\upi}\tau^{3/2}} \exp \left( -{\xi^2+1 \over 2\tau}  \right) \left(\cosh{\xi\over \tau}-\xi \sinh{\xi\over \tau}\right),
\label{eq:n1}
\end{equation}
whereas its Fourier transform is,
\begin{equation}
{\cal G}_{\dot{M}}(\xi,\omega) = {\dot M_0\over 2} {\rm e}^{-(1+{\rm i})(1+\xi) \sqrt{\omega}} \left[1+{\rm e}^{2(1+{\rm i})\xi \sqrt{\omega}}\right]  .
\label{eq:fn1}
\end{equation}

\label{lastpage}
\end{document}